\documentstyle[preprint,aps,tighten]{revtex}

\begin{document}
\draft
\title{Non-linear screening corrections of stellar nuclear reaction rates and their
effects on solar neutrino fluxes}
\author{Theodore E. Liolios \footnote{theoliol@physics.auth.gr}}
\address{Department of Theoretical Physics, University of\\
Thessaloniki, Thessaloniki 54006, Greece}
\date{January 2000}
\maketitle

\begin{abstract}
Non-linear electron-screening corrections of stellar nuclear fusion rates
are calculated analytically in the framework of the Debye-H\"{u}ckel model
and compared with the respective ones of Salpeter's weak screening
approximation. In typical solar conditions, the deviation from Salpeter's
screening factor is less than one percent, while for hotter stars such
corrections turn out to be of the order of one percent only over the limits
of the Debye-H\"{u}ckel model. Moreover, an investigation of the impact of
such non-linear screening effects on the solar neutrino fluxes yields
insignificant corrections for both the $pp$ and $CNO$ chain reactions.
\end{abstract}

\pacs{PACS number(s): 25.10.+s, 25.45.-z, 26.65.+t}

\oddsidemargin -0.25cm \evensidemargin -0.25cm \topmargin -1.0cm \textwidth %
16.3cm \textheight 22.3cm

\section{Introduction}

Stellar nuclear reactions rates are influenced by the electron cloud that
screens fusing nuclei from each other. In recent years, there has been a
wide interest of the actual effect of such screening corrections on reaction
rates. That interest stems mainly from the fact that any variation of the
fusion reaction rates reflects on the solar neutrino fluxes. Therefore the
investigation of the screening effect has produced a number of papers, most
of which come to the conclusion that the solution of the neutrino problem
cannot be found in the screening effect, though it has been argued that the
discrepancy between theory and experiment might be reduced by means of a
suitable screening correction\cite{dar}. Investigations pointing out the
inability of screening corrections to reconcile theory and experiment are
for example that of Ricci et all \cite{ricci} and that of Gruzinov and
Bahcall \cite{gruzbah}. The former modifying Mitler's model\cite{mitler}
showed that possible uncertainties due to screening are too small to solve
the solar neutrino problem while the latter concluded that Salpeter's weak
screening formula is adequate for most solar fusion reactions, thus proving
that screening can only modify solar reaction rates and neutrino fluxes by a
few percent.

However, as there is an exhaustive effort for higher precision in the
theoretical as well as experimental calculation of neutrino fluxes, even
corrections of a few percent have become significant\cite{bah1995}

Most of the studies in that field depart from Salpeter's weak screening
formalism\cite{salpeter}, and consider various corrections such as those
arising from vacuum polarization\cite{gould} , electron density distribution%
\cite{gruzbah} etc. However, even authors who attempt to go beyond the
linear regime make assumptions that lead to inaccuracies. For instance, a
popular oversimplification is to consider that the electron density around
the nucleus is equal to the average electron density in the plasma\cite
{mitler}. In a recent review of solar fusion cross sections\cite{adel}, the
absence of an analytical study of nonlinear screening effects was
underlined, while at the same time it was suggested that electron degeneracy
and non-linearities of the Debye-screening could produce corrections to
Salperer's formula roughly of the order of a few percent for solar reaction.
On the other hand, a numerical integration of the Poisson-Boltzmann equation
for a mixture of electrons and ions \cite{gruzbah}, yielded non-linear
corrections of the order of $1\%.$

The aim of the present paper is to study analytically non-linear (higher
order) corrections for stellar nuclear reactions and determine their effects
on the solar neutrino fluxes. The paper is organized as follows: In unit II
the fundamentals of screened thermonuclear reaction are briefly reviewed.
The non-linear screening formalism is derived in unit III and the effects of
non-linearities on the most probable energy of interaction and the cross
section factor are clarified. In unit IV the derived formula are implemented
for various stellar reactions while in unit V the sensitivity of the solar
neutrino fluxes to such non-linear corrections is investigated. Finally the
conclusions of this paper are given in unit VI. An appendix at the end
elucidates some misconceptions about the effects of electron screening on
the astrophysical factor and the most probable energy of interaction.

\section{Fundamentals of electron screening in thermonuclear reactions}

In the stellar interior the nuclear fusion cross section $\sigma \left(
E\right) $ can be written as a product of a penetration factor $P\left(
E\right) $ times a nuclear factor $\sigma _{nuc}\left( E\right) .$ The
penetration factor is actually the probability that two positively charged
particles will tunnel through the Coulomb barrier that separates them:

\begin{equation}
\sigma \left( E\right) =\sigma _{nuc}\left( E\right) P\left( E\right)
\end{equation}
For two bare nuclei of mass numbers $A_{1},$and $A_{2}$ respectively, that
barrier is given by

\begin{equation}
E_{c}=\frac{Z_{1}Z_{2}e^{2}}{R}\,\,
\end{equation}
where the two nuclei can be considered sharp-edged spheres so that

\begin{equation}
R=1.4\left( A_{1}^{1/3}+A_{2}^{1/3}\right) \,fm  \label{nucradius}
\end{equation}
$A_{1},A_{2}$ being the mass numbers of the fusing nuclei.

If higher relative angular momenta are considered in the collisions then the
height of the barrier is increased by

\begin{equation}
E_{l}=\frac{l\left( l+1\right) \hbar ^{2}}{2\mu r^{2}}
\end{equation}
where $\mu $ is the reduced mass .

In the stellar plasma the kinetic energy of the interacting particles is
determined by a Maxwell-Botzmann distribution of velocities corresponding to
a thermal energy

\begin{equation}
kT=0.086T_{6} \thinspace \thinspace keV
\end{equation}
where $T_{6}$ the temperature in millions degrees Kelvin$.$ $\,$In a
semiclassical approach it can be shown that the two nuclei can only interact
if

\begin{equation}
l\leq 2\cdot 10^{-3}R_{\left( fm\right) }\sqrt{AT_{6}}
\end{equation}
where $A$ is the reduced mass number. Note that all energies are assumed to
be center of mass energies unless specified otherwise. By means of the above
condition it is easy to show that in stellar fusion reactions between light
nuclei $s-$wave interactions dominate. In such a case the interaction is
adequately described by a single particle potential of the form:

\begin{equation}
V\left( r\right) =V_{N}\left( r\right) +V_{c}\left( r\right)
\end{equation}
where $V_{N}\left( r\right) $ is the nuclear potential and $V_{c}\left(
r\right) $ is a potential which describes the Coulomb interaction and is not
necessarily a pure Coulomb potential. Gamow first showed , in connection
with the problem of $\alpha -$decay that for two bare nuclei of charge $%
Z_{1} $ and $Z_{2}$ moving with relative velocity $u$ the probability to
penetrate the Coulomb barrier is proportional to the factor $\exp \left(
-2\pi n\right) $ where $n$ is the Sommerfeld parameter:

\begin{equation}
n\left( E\right) =Z_{1}Z_{2}e^{2}\sqrt{\frac{\mu }{2\hbar ^{2}}}E^{-1/2}
\label{nsom}
\end{equation}
The cross section of the nuclear fusion reaction is then given by:

\begin{equation}
\sigma _{l}\left( E\right) =\frac{S\left( E\right) }{E}T_{l}\left( E\right)
\end{equation}
where

\begin{equation}
T_{l}\left( E\right) =\exp \left[ -\frac{2\sqrt{2\mu }}{\hbar }%
\int_{R}^{r_{c}}\sqrt{V_{c}\left( r\right) +\frac{l\left( l+1\right) }{2\mu
r^{2}}-E}dr\right]
\end{equation}
and $r_{c}$ is the classical turning point given by

\begin{equation}
V\left( r_{c}\right) +\frac{l\left( l+1\right) }{2\mu r_{c}^{2}}=E
\end{equation}
The lower limit $R$ of the integral is the radius of the nuclear forces
given by $\left( \ref{nucradius}\right) $ (assumed to be practically zero)

In the framework of the nearly perfect ionized gas, the presence of the
electron cloud around the nuclei increases the reaction rates over their
laboratory analogs. This screening effects has been studied by many authors
and the most popular potential is the Debye-H\"{u}ckel (D-H) screened
Coulomb potential given by

\begin{equation}
V_{c}\left( r\right) =U_{e}\frac{e^{-u}}{u}  \label{scpotential}
\end{equation}
where $u=r/r_{D}$ and $U_{e}=Z_{1}Z_{2}e^{2}r_{D}^{-1}$ is the screening
potential energy i.e. the Coulomb energy of the two colliding atoms at a
distance equal to the D-H radius $r_{D}$ given by:

\begin{equation}
r_{D}^{-2}=\frac{4\pi e^{2}}{kT}\left( \sum_{i}Z_{i}^{2}n_{i}+n_{e}\theta
_{e}\right)  \label{rd}
\end{equation}
where $n_{i}$ is the number density of ions with charge $Z_{i}$ , $n_{e}\,$%
is the average electron density and $\theta _{e}$ the electron degeneracy
factor.

At this stage we need to define the limits of the Debye-H\"{u}ckel model
which are actually the limits of the results of this paper. This model
assumes a nearly perfect gas at low density where the average Coulomb energy 
$\left\langle E_{c}\right\rangle $ between two adjacent nuclei is much
smaller than the thermal kinetic energy of the plasma

\begin{equation}
\left\langle E_{c}\right\rangle <<kT  \label{ekt}
\end{equation}
Therefore, in the calculations that follow we have to bear in mind that the
derived non-linear corrections cannot be applied beyond the realm of
validity of this condition.

\section{Nonlinear screening formalism}

Let us assume a screening potential of the form $\left( \ref{scpotential}%
\right) .$ The cross section for thermonuclear reactions can be written as :

\begin{equation}
\sigma \left( E\right) =\frac{S\left( E\right) }{E}e^{-4n\left( E\right)
I\left( x\right) }  \label{cs}
\end{equation}
where $x=\frac{r_{c}}{r_{D}}$ .

The quantity $I\left( x\right) $ is

\begin{equation}
I\left( x\right) =e^{-x}\int_{0}^{1}\sqrt{\frac{1}{u}e^{x\left( 1-u\right)
}-1}du
\end{equation}
where $x=x\left( E\right) $ is the solution of the equation:

\begin{equation}
xe^{x}=\frac{U_{e}}{E}  \label{xex}
\end{equation}
and $I\left( 0\right) $ is :

\begin{equation}
I\left( 0\right) =\int_{0}^{1}\sqrt{\frac{1}{u}-1}du=\frac{\pi }{2}
\label{io}
\end{equation}
By introducing a multiplicative corrective term $\xi $ $\left( x\right) $
such that

\begin{equation}
I\left( x\right) =I\left( 0\right) \xi \left( x\right) =\frac{\pi }{2}\xi
\left( x\right)  \label{ioksi}
\end{equation}
we obtain

\begin{equation}
\xi \left( x\right) =e^{-x}\left( 1+\frac{x}{2}+\frac{x^{2}}{16}+\frac{x^{3}%
}{48}-\frac{3x^{4}}{1024}+\frac{13x^{5}}{5120}-\frac{73x^{6}}{49152}%
+...\right)
\end{equation}

The screened reaction rate between nuclei $i$ and $j$ is:

\begin{equation}
r_{ij}^{sc}=\left( 1+\delta _{ij}\right) ^{-1}n_{i}n_{j}\left\langle \sigma
v\right\rangle ^{sc}  \label{rijdelta}
\end{equation}
where $n_{i},n_{j}$ are the number densities of nuclei $i$ and $j\,\,$%
respectively, $\delta _{ij}$ is the Kronecker delta, and the thermalized
cross section per pair of particles is

\begin{equation}
\left\langle \sigma v\right\rangle ^{sc}=\sqrt{\frac{8}{\mu \pi }}\left(
kT\right) ^{-\frac{3}{2}}\int_{0}^{\infty }S\left( E\right) \exp \left[ -%
\frac{E}{kT}-4n\left( E\right) I\left( x\right) \right] dE  \label{thecs}
\end{equation}
Note that in the present paper the superscripts: $nos,\,wes,\,sc,\,ss,\,$%
indicate respectively no-screening, weak-screening, non-linear screening and
strong screening regimes.

Therefore

\begin{equation}
\left\langle \sigma v\right\rangle ^{sc}=\sqrt{\frac{8}{\mu \pi }}\left(
kT\right) ^{-\frac{3}{2}}f_{0}\left( E_{0}^{sc}\right) \int_{0}^{\infty
}S\left( E\right) \exp \left[ -\frac{E}{kT}-2\pi n\left( E\right) \right] dE
\label{sigmaf}
\end{equation}
where the screening correction factor $f_{0}$ which is now evaluated at the
most probable energy of the screened interaction is given by:

\begin{equation}
\ln f_{0}=\pi n\left( E_{0}^{sc}\right) \left( x-\frac{x^{2}}{8}-\frac{x^{3}%
}{12}+\frac{35x^{4}}{512}-\frac{23x^{5}}{640}+\frac{449x^{6}}{24576}+O\left[
x^{7}\right] \right)  \label{fsc}
\end{equation}
Following the method of the steepest decent\cite{clayton}, the most
effective energy of interaction $E_{0}^{sc}$ between the two screened nuclei
is:

\begin{equation}
\frac{d}{dE}\left( \frac{E}{kT}+4n\left( E\right) I\left( x\right) \right)
_{E=E_{0}^{sc}}=0
\end{equation}
which yields (see Appendix I):

\begin{equation}
E_{0}^{sc}=\left( 2I\left( x\right) Z_{1}Z_{2}e^{2}\sqrt{\frac{\mu }{2\hbar
^{2}}}kT\right) ^{\frac{2}{3}}  \label{escix}
\end{equation}
where $x=x\left( E_{0}^{sc}\right) $ is the solution of equation $\left( \ref
{xex}\right) .$

Hence

\begin{equation}
E_{0}^{sc}=\xi ^{2/3}\left( x\right) E_{0}^{nos}  \label{escnos}
\end{equation}

On the other hand by using the formula for the most effective energy of
interaction for two bare nuclei we obtain:

\begin{equation}
E_{0}^{sc}=1.220\cdot \left( Z_{1}^{2}Z_{2}^{2}AT_{6}^{2}\right) ^{1/3}\xi
^{2/3}\left( x\right) \,\,keV
\end{equation}
Eq.$\left( \ref{xex}\right) $ now reads:

\begin{equation}
xe^{x}\xi ^{2/3}\left( x\right) =\frac{U_{e}}{E_{0}^{nos}}=\frac{1180\left(
Z_{1}Z_{2}\right) ^{1/3}}{\left( AT_{6}^{2}\right) ^{1/3}\,r_{D\left(
fm\right) }}  \label{hixi}
\end{equation}
In the weak screening $\left( wes\right) $ approximation, to first order in
x, we obtain:

\begin{equation}
E_{0}^{wes}\simeq 1.220\cdot \left( Z_{1}^{2}Z_{2}^{2}AT_{6}^{2}\right)
^{1/3}\left( 1-\frac{x}{3}\right)
\end{equation}
which corrects the result of ref $\cite{error}$ according to which the
energy should be shifted by $\left( 1-x\right) ,$ instead.

Moreover, in the weak screening approximation, the integral $I\left(
x\right) $ can be written:

\begin{equation}
I^{wes}\left( x\right) \simeq \frac{\pi }{2}\left( 1-\frac{x}{2}\right)
\end{equation}
and therefore one obtains Salpeter's weak screening correction\cite{error}:

\begin{equation}
f_{0}^{wes}\simeq e^{\pi nx}
\end{equation}
Note that the assumption made in ref\cite{error} about the classical turning
point being equal for both the screened and unscreened cases is not
necessary \cite{chen}. In fact the actual approximation is

\begin{equation}
r_{c}^{sc}\simeq r_{c}^{nos}\left( 1-x\right)
\end{equation}
Regarding the rate of thermonuclear reactions , it has been shown that in
the stellar interior where the temperature is $T$ degrees Kelvin and the
density is $\rho $ in $g/cm^{3}$ the nuclear reaction rate is:$.$%
\begin{equation}
r_{ij}=\frac{2.62\cdot 10^{29}}{\left( 1+\delta _{ij}\right) }\rho ^{2}\frac{%
X_{i}X_{j}}{A_{i}A_{j}Z_{i}Z_{j}}f_{0}S_{eff}\tau ^{2}e^{-\tau
}\,\,cm^{-3}\sec ^{-1}  \label{rij}
\end{equation}
where

\begin{equation}
\tau =\frac{3E_{0}^{nos}}{kT}
\end{equation}
and $S_{eff}$ is expressed in keV-barns. The quantities $%
X_{i,j},Z_{i,j},A_{i,j},$are the mass fraction, charge and mass number
respectively of the nucleus $i,j$ .

To first order in $\tau ^{-1}\cite{seff}$ :

\begin{equation}
S_{eff}\left( E_{0}\right) \simeq S\left( 0\right) \left[ 1+\frac{5}{12\tau }%
+S^{-1}\left( 0\right) \left( \frac{dS}{dE}\right) _{E=0}\left( E_{0}+\frac{%
35}{36}kT\right) +...\right]
\end{equation}
\newline

Although in Eq.$\left( \ref{rij}\right) $ the screening enhancement factor $%
f_{0}$ is now evaluated at $E_{0}^{sc}$ the effective astrophysical factor $%
S_{eff}$ is always evaluated at $E_{0}^{nos}$ as the screening effects have
already been worked out of the integral so the correction assumed in ref.%
\cite{error} is not applicable (see Appendix II)

\section{Results for various stellar fusion reactions}

In the solar region of maximum energy production $\left( R/R_{\odot
}=0.09\right) ,$ the temperature and the density are respectively $%
T_{6}=13.5 $ and $\rho =93.3\frac{g}{cm^{3}}$. Using recently calculated
isotopic abundances of the solar interior\cite{seismo} the average
internuclear distance given by

\begin{equation}
a=\frac{1}{\left( 4\pi \rho N_{0}\right) ^{1/3}}=51000\rho ^{-1/3}fm
\label{avdist}
\end{equation}
is found to be $a\simeq 11244\,fm$ and the Debye-Huckel radius is $%
r_{D}=25719\,fm$, where the weak electron degeneracy has been taken into
account $\left( \theta _{e}\simeq 0.92\right) .$ Moreover, the thermal
kinetic energy is $kT=1.161$ $keV$ and is always higher than the average
Coulomb energy between two adjacent ions according to condition $\left( \ref
{ekt}\right) $. Therefore the potential of \ Eq.$\left( \ref{scpotential}%
\right) $ can be used in order to study the screening corrections to solar
fusion reactions. The results are depicted in Table 1 that follows:

\medskip \medskip

\begin{center}
{\bf Table I. The electron screening corrections for various solar fusion
reactions.}{\small \ $
\begin{array}{lllllllll}
Fusion\,\,reaction & f_{0}^{wes} & f^{sc} & \Delta f^{a} & \Delta f^{n} & 
E_{0}^{nos}\left( keV\right) & E_{0}^{sc}\left( keV\right) & E_{c}\left(
keV\right) & U_{e}\left( keV\right) \\ 
H^{1}\left( p,e^{+}\nu _{e}\right) H^{2} & 1.049 & 1.048 & 0.030 & 0.5 & 
5.624 & 5.471 & 514 & 0.056 \\ 
He^{3}\left( _{2}He^{4},\gamma \right) Be^{7} & 1.212 & 1.210 & 0.129 & 1.7
& 20.860 & 20.786 & 1358 & 0.223 \\ 
He^{3}\left( _{2}He^{3},2p\right) He^{4} & 1.212 & 1.210 & 0.135 & - & 19.952
& 19.877 & 1426 & 0.224 \\ 
Be^{7}\left( p,\gamma \right) B^{8} & 1.212 & 1.210 & 0.161 & 1.5 & 16.671 & 
16.596 & 1412 & 0.224 \\ 
N^{14}\left( p,\gamma \right) O^{15} & 1.400 & 1.396 & 0.332 & 0.8 & 24.735
& 24.606 & 2111 & 0.391 \\ 
C^{12}\left( p,\gamma \right) N^{13} & 1.335 & 1.331 & 0.272 & - & 22.238 & 
22.127 & 1876 & 0.348
\end{array}
$ }
\end{center}

{\small Table I. The electron screening corrections for various solar fusion
reactions with the following notation : }$f_{0}^{wes},f_{0}^{sc},${\small %
the screening factors for the weak and non-linear screening respectively, }$%
\Delta f^{a}\left( \%\right) ${\small \ the non-linear correction of the
weak screening in percent, }$E_{0}^{nos},E_{0}^{sc}${\small \ the most
probable energy of interaction for the two cases, }$E_{c}${\small \ the
coulomb barrier of the reaction and }$U_{e}${\small \ coulomb energy of two
ions at distance }$r_{D}.${\small \ Moreover the available corrections of
the numerical approach\cite{gruzbah} are also shown as }$\Delta f^{n}\left(
\%\right) .$

\medskip \medskip

Note that the statement\cite{adel} that non-linearities of the Debye
screening might produce a correction to Salpeter's formula of the order of a
few percent can be now upgraded. According to the above results, as far as
the Debye-H\"{u}ckel mean field potential is concerned, the actual deviation
is less than $0.5\%$ for all solar fusion reactions.

A very interesting fact is that such non-linear corrections are largely due
to the presence of Eq.$\left( \ref{hixi}\right) $ which supersedes its
linear oversimplified counterpart of the weak screening regime:

\begin{equation}
x=\frac{U_{e}}{E_{0}^{nos}}
\end{equation}
That means that for most stellar conditions instead using Eq.$\left( \ref
{fsc}\right) $ we can safely neglect higher order terms in order to use:

\begin{equation}
\ln f_{0}=\pi n\left( E_{0}^{sc}\right) x\left( E_{0}^{sc}\right)
\end{equation}
where $x\left( E_{0}^{sc}\right) $ is always the solution of Eq.$\left( \ref
{hixi}\right) $

Moreover, the usual assumption\cite{ricci} that the screening enhancement
factors are independent of the isotope is not necessarily valid when
nonlinear corrections are considered. This is also clear from Eq.$\left( \ref
{hixi}\right) $ where $x$ depends on the reduced mass number. However, as
can be readily seen from table I, for typical solar conditions, isotopic
dependence is indeed practically negligible. On the other hand the Gamow
peak energy, which is usually considered screening independent, in the
non-linear regime shows a maximum variation of the order of $0.5\%\,$.

For purposes of illustration nonlinear screening corrections have also been
calculated for the cases considered by Salpeter\cite{salpeter}. Following
the notation of table I we have obtained the following results:

\medskip

a) In a red dwarf (main sequence star cooler then the sun), with typical
central conditions $\rho =100,\,T_{6}=8,\theta _{e}=0.8,$ the screening
corrections for the $pp\,$reaction are:

\begin{equation}
\begin{array}{llllllll}
f_{0}^{wes} & f^{sc} & \Delta f\left( \%\right) & E_{0}^{nos}\left(
keV\right) & E_{0}^{sc}\left( keV\right) & E_{c}\left( keV\right) & 
U_{e}\left( keV\right) & kT\left( keV\right) \\ 
1.116 & 1.114 & 0.132 & 3.873 & 3.848 & 514 & 0.076 & 0.688
\end{array}
\end{equation}

b) Salpeter also considers the reaction $N^{14}\left( p,\gamma \right)
O^{15} $ for the interior of the hotter main sequence stars for a
combination: $\rho =122,\,T_{6}=11.6,\theta _{e}=0.85.\,$By taking into
account non-linear corrections the results are:

\begin{equation}
\begin{array}{llllllll}
f_{0}^{wes} & f^{sc} & \Delta f\left( \%\right) & E_{0}^{nos}\left(
keV\right) & E_{0}^{sc}\left( keV\right) & E_{c}\left( keV\right) & 
U_{e}\left( keV\right) & kT\left( keV\right) \\ 
1.648 & 1.636 & 0.685 & 22.356 & 22.193 & 2111 & 0.499 & 0.997
\end{array}
\end{equation}
For the same reaction in the center of Sirius where $\rho =80$ , $T_{6}=20,$%
and $\theta _{e}\simeq 1.$ the corrections are:

\begin{equation}
\begin{array}{llllllll}
f_{0}^{wes} & f^{sc} & \Delta f\left( \%\right) & E_{0}^{nos}\left(
keV\right) & E_{0}^{sc}\left( keV\right) & E_{c}\left( keV\right) & 
U_{e}\left( keV\right) & kT\left( keV\right) \\ 
1.209 & 1.207 & 0.120 & 32.145 & 32.037 & 2111 & 0.327 & 1.720
\end{array}
\end{equation}

c) In the more luminous main sequence stars where Hydrogen has been
exhausted, the first stage of the two-stage reaction $3He^{4}\longrightarrow
C^{12}+\gamma $ is the temporary formation of $Be^{8},$through the reaction $%
He^{4}+He^{4}\longrightarrow B^{8}$. The screening corrections for such a
collision between two alpha particles, for indicative conditions considered
by Salpeter are:

i) $\rho =10^{4},\,T_{6}=150,\theta _{e}=0.87$

\begin{equation}
\begin{array}{llllllll}
f_{0}^{wes} & f^{sc} & \Delta f\left( \%\right) & E_{0}^{nos}\left(
keV\right) & E_{0}^{sc}\left( keV\right) & E_{c}\left( keV\right) & 
U_{e}\left( keV\right) & kT\left( keV\right) \\ 
1.049 & 1.048 & 0.017 & 109.346 & 109.141 & 1295 & 0.619 & 12.9
\end{array}
\end{equation}

Despite the high density , obviously condition $\left( \ref{ekt}\right) $
still holds here , therefore Salpeter's weak screening approximation is
still reliable.

ii) $\rho =10^{6},\,T_{6}=150,\theta _{e}\simeq 0$

\begin{equation}
\begin{array}{llllllll}
f_{0}^{wes} & f^{sc} & \Delta f\left( \%\right) & E_{0}^{nos}\left(
keV\right) & E_{0}^{sc}\left( keV\right) & E_{c}\left( keV\right) & 
U_{e}\left( keV\right) & kT\left( keV\right) \\ 
1.491 & 1.474 & 1.136 & 109.346 & 107.680 & 1295 & 5.540 & 12.9
\end{array}
\end{equation}
In the above case $\left( ii\right) ,$ the electron screening effect is
strong enough for the weak screening approximation to become inaccurate and
therefore nonlinear corrections become important. Unfortunately, even
nonlinear corrections cannot redeem the inaccuracies of the Debye-H\"{u}ckel
potential which is only applicable to fusion reactions where condition $%
\left( \ref{ekt}\right) $ holds. On the other hand the strong screening
formula\cite{salpeter}

\begin{equation}
f_{o}^{ss}=\exp \left[ 0.205\left( \frac{\rho }{\mu _{e}}\right)
^{1/3}T_{6}^{-1}\left[ \left( Z_{1}+Z_{2}\right)
^{5/3}-Z_{1}^{5/3}-Z_{2}^{5/3}\right] \right]  \label{fss}
\end{equation}
where $\mu _{e}$ is the mean molecular weight per electron, can be
cautiously used here as screening is strong, though not strong enough to
assume absolute validity . By applying formula $\left( \ref{fss}\right) $ we
obtain $f_{0}^{ss}=1.498$ which is reasonably close to its linear and
non-linear counterparts.

\section{Effects on solar neutrino fluxes}

It has been suggested\cite{shaviv} that the electron screening effect may be
important to the solar neutrino production as it enhances the reaction rates
in the interior of the sun. More precisely, if we consider the Gamow peak
screening-independent then any variations of screening reflects on the cross
section $S\left( E\right) $ which is in fact multiplied by this screening
factor so that:

\begin{equation}
S^{sc}\left( E\right) =f_{0}S^{nos}\left( E\right)  \label{sfs}
\end{equation}
In fact, it has been shown that the solar neutrino fluxes $\Phi $ can be
given as a function of the screening factors of the $pp$ and $CNO$ chains%
\cite{ricci}. In that work the screening factor $f$ was assumed to be
isotope independent. Although in typical solar conditions this assumption
was shown to be valid (see table I) when nonlinear screening effects are
considered the derived formula should be modified as follows:

$Be^{7}\left( e^{-},\nu _{e}\right) Li^{7}:$

\begin{equation}
\Phi _{Be}^{sc}=\Phi _{Be}^{nos}\left( f_{p+p}\right) ^{-10/8}\frac{%
f_{He^{3}+He^{4}}}{\sqrt{f_{He^{3}+He^{3}}}}  \label{fbe}
\end{equation}

$Be^{7}\left( p,\gamma \right) B^{8}\left( e^{+},\nu _{e}\right) B^{8*}:$

\begin{equation}
\Phi _{B}^{sc}=\Phi _{B}^{nos}\left( f_{p+p}\right) ^{-23.6/8}\left(
f_{p+^{7}Be}\right) \frac{f_{He^{3}+He^{4}}}{\sqrt{f_{He^{3}+He^{3}}}}
\label{fb}
\end{equation}

On the other hand for the $CNO$\ cycle which is governed by the slowest
reaction $N^{14}\left( p,\gamma \right) O^{15}$ the enhancement on the
neutrino fluxes are\cite{ricci}:

$N^{13}\left( e^{+}\nu _{e}\right) C^{13}$ and $O^{15}\left( e^{+},\nu
_{e}\right) N^{15}:$

\begin{equation}
\Phi _{N,O}^{sc}=\Phi _{N,O}^{nos}\left( f_{p+p}\right) ^{-22/8}\left(
f_{p+^{14}N}\right)  \label{fno}
\end{equation}
Conservation of luminosity yields the pp neutrino fluxes by means of

\begin{equation}
\Phi _{pp}^{sc}+\Phi _{Be}^{sc}+\Phi _{N}^{sc}+\Phi _{O}^{sc}=\Phi
_{pp}^{nos}+\Phi _{Be}^{nos}+\Phi _{N}^{nos}+\Phi _{O}^{nos}  \label{ffff}
\end{equation}
whereas the $pep$ neutrino fluxes can be obtained by the observation that in
any solar model:

\begin{equation}
\frac{\Phi _{pep}}{\Phi _{pep}^{nos}}=\frac{\Phi _{pp}}{\Phi _{pp}^{nos}}
\label{ppep}
\end{equation}
Instead of solving algebraic equations, an easier way of obtaining
corrections to $pp$ and $pep$ neutrino fluxes is by using the
proportionality formula\cite{neuastro}:

\begin{equation}
\Phi _{pp} \sim
S_{pp}^{0.14}S_{He^{3}+He^{3}}^{0.03}S_{He^{3}+He^{4}}^{-0.06}  \label{fpp}
\end{equation}
which readily yields:

\begin{equation}
\frac{\Phi _{pp}^{sc}}{\Phi _{pp}^{nos}}
=f_{p+p}^{0.14}f_{He^{3}+He^{3}}^{0.03}f_{He^{3}+He^{4}}^{-0.06}
\label{fbah}
\end{equation}
By using the above formulae the corrections to the solar neutrino fluxes
both in the weak screening and non-linear screening regimes are shown in
table II.

\newpage

{\bf Table II. The nonlinear screening corrections of solar neutrino fluxes}

\[
\begin{array}{lll}
\,Neutrino\,\,\,source & \Phi ^{wes}/\Phi ^{nos} & \Phi ^{sc}/\Phi ^{nos} \\ 
H^{1}\left( p,e^{+}\nu _{e}\right) H^{2} & 1.000 & 1.000 \\ 
H^{1}\left( pe^{-},\nu _{e}\right) H^{2} & 1.001 & 1.001 \\ 
Be^{7}\left( e^{-},\nu _{e}\right) Li^{7} & 1.037 & 1.037 \\ 
B^{8}\left( e^{+},\nu _{e}\right) B^{8*} & 1.158 & 1.159 \\ 
N^{13}\left( e^{+}\nu \right) C^{13} & 1.227 & 1.226 \\ 
O^{15}\left( e^{+},\nu _{e}\right) N^{15} & 1.227 & 1.226
\end{array}
\]

Admittedly, it has been argued\cite{dzitko} that the weak-screening factors
are more appropriate for the $pp$ reaction rate where condition $\left( \ref
{ekt}\right) \,$is fully satisfied, whereas for reactions with $Z_{1}Z_{2}>4$
Mitler's formula\cite{mitler} should be used, instead. In our approach
(Table II) we have used the weak-screening formula for all solar reactions
following the recent suggestion of ref.\cite{gruzbah}. Note that in ref.\cite
{ricci}, where Mitler's screening factors were used, the ratio $\Phi
^{sc}/\Phi ^{nos}$ for the $pp\,$neutrino was found to be $0.995$ while for
the $N^{13\text{ }}$neutrino it was $1.13$. This discrepancy is unimportant
to the results of this paper according to which there is a negligible
contribution of the investigated non-linear screening effects to solar
neutrino fluxes well below the experimental errors of any currently
imaginable neutrino detector.

\section{Conclusions}

In this paper we have studied the non-linear effects of electron screening
on stellar nuclear fusion rates calculating the respective corrections
analytically. The formalism employed has been based on the Debye-H\"{u}ckel
model. In typical solar conditions such non-linear effects are shown to be
negligible proving Salpeter's linear approach to be sufficient for the study
of solar nuclear reactions. Regarding the solar neutrino problem it was also
shown that, non-linear screening leads to a negligible correction of the
solar neutrino fluxes of the $pp$ and $CNO$ chains. Moreover, non-linear
corrections are shown to be of some importance only in the intermediate
screening regime, where the average Coulomb energy begins to challenge the
average thermal kinetic energy and the Debye-H\"{u}ckel model begins to
break down.

\medskip

ACKNOWLEDGMENTS

This work was financially supported by the Greek State Grants Foundation
(IKY)

\medskip

{\bf APPENDIX}

{\bf I. The most effective energy of interaction.}

The method of the steepest decent yields the maximum of the integrand of Eq.$%
\left( \ref{thecs}\right) $:

\begin{equation}
\frac{d}{dE}\left( \frac{E}{kT}+4n\left( E\right) I\left( x\right) \right)
_{E=E_{0}^{sc}}=0
\end{equation}
In fact the quantity $x$ is energy dependent $x=x\left( E\right) $ by means
of Eq..$\left( \ref{xex}\right) $, therefore upon differentiation we obtain:%
\newline
\begin{equation}
\frac{1}{kT}+2\pi \left[ \left( \frac{dn\left( E\right) }{dE}\right)
_{E=E_{0}^{sc}}\xi \left( x_{0}^{sc}\right) +n\left( E_{0}^{sc}\right)
\left( \frac{dx\left( E\right) }{dE}\right) _{E=E_{0}^{sc}}\left( \frac{d\xi
\left( x\right) }{dx}\right) _{x_{0}^{sc}}\right] =0
\end{equation}
where $n\left( E\right) $ is given by $\left( \ref{nsom}\right) $ and $%
x_{0}^{sc}$ is the solution of Eq..$\left( \ref{xex}\right) .\,$If we assume
that throughout the integral which appears in the thermalized cross section$%
\,$in Eq.$\left( \ref{thecs}\right) $ there is only a negligible variation
of $\xi \left( x\right) $ then the above integral yields Eq.$\left( \ref
{escix}\right) .\,$This assumption is better than assuming that $\xi \left(
x\right) =1,$ which practically yields $E_{0}^{nos}=E_{0}^{sc},$ as the
latter disregards {\it a priori} all the corrections of the screened
interaction to the effective energy of interaction .

{\bf II. Screening independence of $S_{eff}$ }.

The basic formula for the thermalized cross section of the non-resonant
screened thermonuclear reaction is : 
\begin{equation}
\left\langle \sigma v\right\rangle ^{sc}=\sqrt{\frac{8}{\mu \pi }}\left(
kT\right) ^{-\frac{3}{2}}\int_{0}^{\infty }S\left( E\right) \exp \left[ -%
\frac{E}{kT}-4n\left( E\right) I\left( x\right) \right] dE
\end{equation}
We have to work out the screening factor $f_{0}\,$first otherwise the
introduction of an $S_{eff}^{sc}$ for the screened case replaces the
integral above by an average (corrected) expression as follows\cite{seff}:

\begin{equation}
\int_{0}^{\infty }S\left( E\right) \exp \left[ -\frac{E}{kT}-4n\left(
E\right) I\left( x\right) \right] dE=2E_{0}^{sc}\left( \frac{\pi }{\tau }%
\right) ^{1/2}e^{-\tau }S_{eff}\left( E_{0}^{sc}\right)
\end{equation}
Once the integral has been replaced the screening effect cannot be
parametrized in the usual way by means of the screening factor $f_{0}.$

Therefore, after we work out the screening enhancement factor $f_{0}$ we
obtain: 
\begin{equation}
\left\langle \sigma v\right\rangle ^{sc}=\sqrt{\frac{8}{\mu \pi }}\left(
kT\right) ^{-\frac{3}{2}}f_{0}\left( E_{0}^{sc}\right) \int_{0}^{\infty
}S\left( E\right) \exp \left[ -\frac{E}{kT}-2\pi n\left( E\right) \right] dE
\end{equation}
where it should be noted that the evaluation of $f_{0}\,$is performed at the
most probable energy of the screened interaction. The remaining integral is
actually unaware of the screening effects as the former is calculated by the
method of the steepest decent around its maximum value, which is the
(unscreened) most effective energy of interaction $E_{0}^{nos}.$

Therefore, the energy $E_{0}$ appearing in the formula:

\begin{equation}
S_{eff}=S\left( E_{0}\right) \left\{ 1+\tau ^{-1}\left[ \frac{5}{12}+\frac{5%
}{2}\frac{S^{^{\prime }}\left( E_{0}\right) E_{0}}{S\left( E_{0}\right) }+%
\frac{S^{^{\prime \prime }}\left( E_{0}\right) E_{0}^{2}}{S\left(
E_{0}\right) }\right] +O\left[ \tau ^{-2}\right] \right\}
\end{equation}
can only be the quantity $E_{0}^{nos}=1.220\cdot \left(
Z_{1}^{2}Z_{2}^{2}AT_{6}^{2}\right) ^{1/3}$, which corresponds to the
no-screening regime. As a result, no screening correction can be
incorporated into the frequently used formula:

\begin{equation}
S_{eff}\simeq S\left( 0\right) \left\{ 1+\frac{5}{12\tau }+\frac{S^{^{\prime
}}\left( 0\right) }{S\left( 0\right) }\left( E_{0}+\frac{35}{36}kT\right) +%
\frac{S^{^{\prime \prime }}\left( 0\right) }{S\left( 0\right) }E_{0}\left( 
\frac{E_{0}}{2}+\frac{89}{72}kT\right) \right\}
\end{equation}
because $E_{0}$ cannot be replaced by $E_{0}^{sc}$. Note that even if that
was the case one should have to modify $\tau =\tau \left( E_{0}^{nos}\right) 
$ as well.

\end{document}